\documentclass[
,twocolumn,twocolappendix]{aastex631}
\hypersetup{linkcolor=blue,citecolor=blue,filecolor=blue,urlcolor=blue}
\usepackage{graphicx}
\usepackage{amssymb}

\shorttitle{Spectra of Solid HCN for Comparison with JWST Data}
\shortauthors{Ozhiganov M. et al.}
\begin{document}
\title{Infrared Spectra of Solid HCN Embedded in Various Molecular Environments for Comparison with the Data Obtained with JWST}

\correspondingauthor{Gleb Fedoseev}
\email{g.s.fedoseev@urfu.ru}

\author[0000-0002-7308-9056]{Maksim Ozhiganov}
\affiliation{Research Laboratory for Astrochemistry, \\ Ural Federal University, Kuibysheva St. 48 \\
Yekaterinburg 620026, Russia}
\author[0009-0001-3921-604X]{Mikhail Medvedev}
\affiliation{Research Laboratory for Astrochemistry, \\ Ural Federal University, Kuibysheva St. 48 \\
Yekaterinburg 620026, Russia}
\author[0009-0007-0109-3439]{Varvara Karteyeva}
\affiliation{Research Laboratory for Astrochemistry, \\ Ural Federal University, Kuibysheva St. 48 \\
Yekaterinburg 620026, Russia}
\author[0009-0004-5420-8824]{Ruslan Nakibov}
\affiliation{Research Laboratory for Astrochemistry, \\ Ural Federal University, Kuibysheva St. 48 \\
Yekaterinburg 620026, Russia}
\author[0009-0008-4084-225X] {Uliana	Sapunova}
\affiliation{Research Laboratory for Astrochemistry, \\ Ural Federal University, Kuibysheva St. 48 \\
Yekaterinburg 620026, Russia}
\author[0000-0001-9388-691X]{Vadim Krushinsky}
\affiliation{Research Laboratory for Astrochemistry, \\ Ural Federal University, Kuibysheva St. 48 \\
Yekaterinburg 620026, Russia}
\author[0009-0005-7855-1923]{Ksenia Stepanova}
\affiliation{Research Laboratory for Astrochemistry, \\ Ural Federal University, Kuibysheva St. 48 \\
Yekaterinburg 620026, Russia}
\author[0009-0006-0032-6707]{Anastasia Tryastsina}
\affiliation{Research Laboratory for Astrochemistry, \\ Ural Federal University, Kuibysheva St. 48 \\
Yekaterinburg 620026, Russia}
\affiliation{Chemistry Department,\\ Lomonosov Moscow State University, Leninskie Gory 1, bld. 3, \\  Moscow 119991, Russia}
\author[0000-0001-8758-9046]{Aleksandr Gorkovenko}
\affiliation{Solid State Magnetism Department, \\ Ural Federal University, Kuibysheva St. 48 \\
Yekaterinburg 620026, Russia}
\author[0000-0003-2434-2219]{Gleb Fedoseev}
\affiliation{Research Laboratory for Astrochemistry, \\ Ural Federal University, Kuibysheva St. 48 \\
Yekaterinburg 620026, Russia}
\author[0000-0003-1684-3355]{Anton Vasyunin}
\affiliation{Research Laboratory for Astrochemistry, \\ Ural Federal University, Kuibysheva St. 48 \\
Yekaterinburg 620026, Russia}
\begin{abstract}

HCN molecules serve as an important tracer for chemical evolution of elemental nitrogen in the regions of star and planet formation. This is largely explained by the fact that N atoms and N$_2$ molecules are poorly accessible for the observation in the radio and infrared ranges. In turn, gas-phase HCN can be observed at various stages of star formation including disks arounds young stars, cometary comas and atmospheres of the planetary satellites. Despite the large geography of gas-phase observations, an identification of interstellar HCN ice is still lacking. In this work we present a series of infrared spectroscopic measurements performed at the new ultra-high vacuum cryogenic apparatus aiming to facilitate the search for interstellar HCN ice. A series of high resolution laboratory infrared spectra of HCN molecules embedded in the H$_2$O, H$_2$O:NH$_3$, CO, CO$_2$ and CH$_3$OH ices at 10~K temperature is obtained. These interstellar ice analogues aim to simulate the surroundings of HCN molecules by the main constituents of the icy mantles on the surface of the interstellar grains. In addition, the spectra of HCN molecules embedded in the solid C$_6$H$_6$, C$_5$H$_5$N and C$_6$H$_5$NH$_2$ are obtained to somehow simulate the interaction of HCN molecules with carbonaceous material of the grains rich in polycyclic aromatic hydrocarbons. The acquired laboratory spectroscopic data are compared with the publicly available results of NIRSpec James Webb Space Telescope observations towards quiescent molecular clouds performed by the ICEAge team.

\end{abstract}


\section{Introduction} \label{sec:introduction}

Hydrogen cyanide (HCN) is one of the first nitrogen-bearing molecules observed in the interstellar medium (ISM), see \cite{Snyder_&_Buhl_1971}. HCN molecules are present in abundance in the gas phase in various types of astronomical objects associated with the different stages of the star formation process. These include diffuse clouds (e.g., \cite{Liszt_&_Lucas_2001}), translucent clouds (e.g., \cite{Hogerheijde_et_al._1995}), dark clouds (e.g., \cite{1992IAUS..150..171O}), low-mass protostars (e.g., \cite{refId0})  and protoplanetary disks (e.g., \cite{g}). Ubiquity  of HCN molecules makes its emission an important tracer for physical and chemical processes occurring during the star formation process, see, for example, \cite{Graninger2014ApJ...787...74G}. Due to the direct chemical links between HCN and its isomer HNC as well as with CN radicals, HCN emission is also suggested as a probe for the protoplanetary disks structure and temperature \citep{Long_et_al._2021}. 

Recently, HCN was detected in the disk around low-mass (0.14~M$\odot$) star 2MASS-J16053215-1933159 using James Webb Space Telescope (JWST) through the emission at 14~$\mu$m with the maximum column density as high as 1.5~\texttimes~10$^{17}$~cm$^{-2}$ \citep{Tabone_et_al._2023,van_Dishoeck_2023}. HCN is detected along with nearly equally abundant C$_6$H$_6$ (benzene), C$_4$H$_2$ (diacetylene) and considerably more abundant C$_2$H$_2$ (acetylene), while the lack of NH$_3$ detection in this source makes HCN the main observable nitrogen reservoir. \cite{Tabone_et_al._2023} propose the existence of a “soot line” in the disk around 2MASS-J16053215-1933159 as one of the possible explanations for the high abundance of the observed hydrocarbons. The “soot line” is associated with the temperature at which refractory carbonaceous material of the grains erodes and sublimates enriching the gas phase with hydrocarbons. Simultaneous detection of HCN along with the hydrocarbons hints for the possible origin of HCN from the solid state, either bonded to the low volatile carbonaceous species or produced by the erosion of N-heterocyclic polyaromatic hydrocarbons. 

Nevertheless, all the previous attempts to identify HCN in the solid state toward various sources using observations in the mid-infrared range have been unsuccessful \citep{Boogert,McClure_et_al._2023}. This can be partially explained by the lack of the available laboratory spectroscopic data for HCN molecules in various chemical environment. Recent results of \cite{Brunken_2024} devoted to the laboratory and astronomical investigation of $\nu_3$ absorption feature of $^{13}$CO$_2$ ice reveal the dependency between the molecular environment surrounding $^{13}$CO$_2$ molecules and the shape and position of their $\nu_3$ absorption feature. Similar effect can be expected for the HCN molecules embedded in various molecular environments. This effect can be particularly prominent for the molecules which cannot form hydrogen bonds with HCN, e.g., CO$_2$ or CO. Previous laboratory studies are focused on the acquisition of mid-infrared spectra of pure HCN ice, which existence in the real astrochemical environments can be doubted, or mixed ices of HCN with predominance of H$_2$O and NH$_3$ \citep{Gerakines_Moore_&_Hudson_(2004),10.1093/mnras/sts272,Gerakines_Yarnal_and_Hudson_(2022)}. Both H$_2$O and NH$_3$ environments are carbon-poor environments. Even though HCN molecules contain a carbon atom and there are laboratory data demonstrating formation of HCN by energetic processing of various carbon-rich ices \citep{STRAZZULLA200013,MOORE2003486,Fedoseev2018}, the systematic measurements of the profile and optical depth of HCN absorption features in carbon-rich environments are missing up to date. 

In this study the mid-infrared spectra of pure HCN ice and the ice mixtures of HCN with H$_2$O and NH$_3$ are obtained for a comparison with the previous works. Then the spectroscopic data for the HCN molecules embedded in CO, CO$_2$, CH$_3$OH (methanol), C$_6$H$_6$, C$_5$H$_5$N (pyridine) and C$_6$H$_5$NH$_2$ (aniline) ices are acquired. The focus is made on the acquisition of data for the $\nu_3$ and $\nu_1$ vibration modes of HCN accessible for the observations with the NIRSpec instrument of JWST. The $\nu_1$ absorption feature of HCN is characterized by the highest optical depth, see \cite{Gerakines_Yarnal_and_Hudson_(2022)}, while the choice of $\nu_3$ absorption feature is motivated by the lack of other abundant interstellar ice constituents exhibiting absorption in the same range \citep{Boogert, McClure_et_al._2023}. All key experiments are performed under the identical conditions with the identical deposition rate of HCN. This allows for the direct comparison of the HCN optical depths in various environments and acquisition of the relative band strength values.

\section{Experimental} \label{sec:experimental}

\begin{figure*}[ht!]
\includegraphics[scale=0.35]{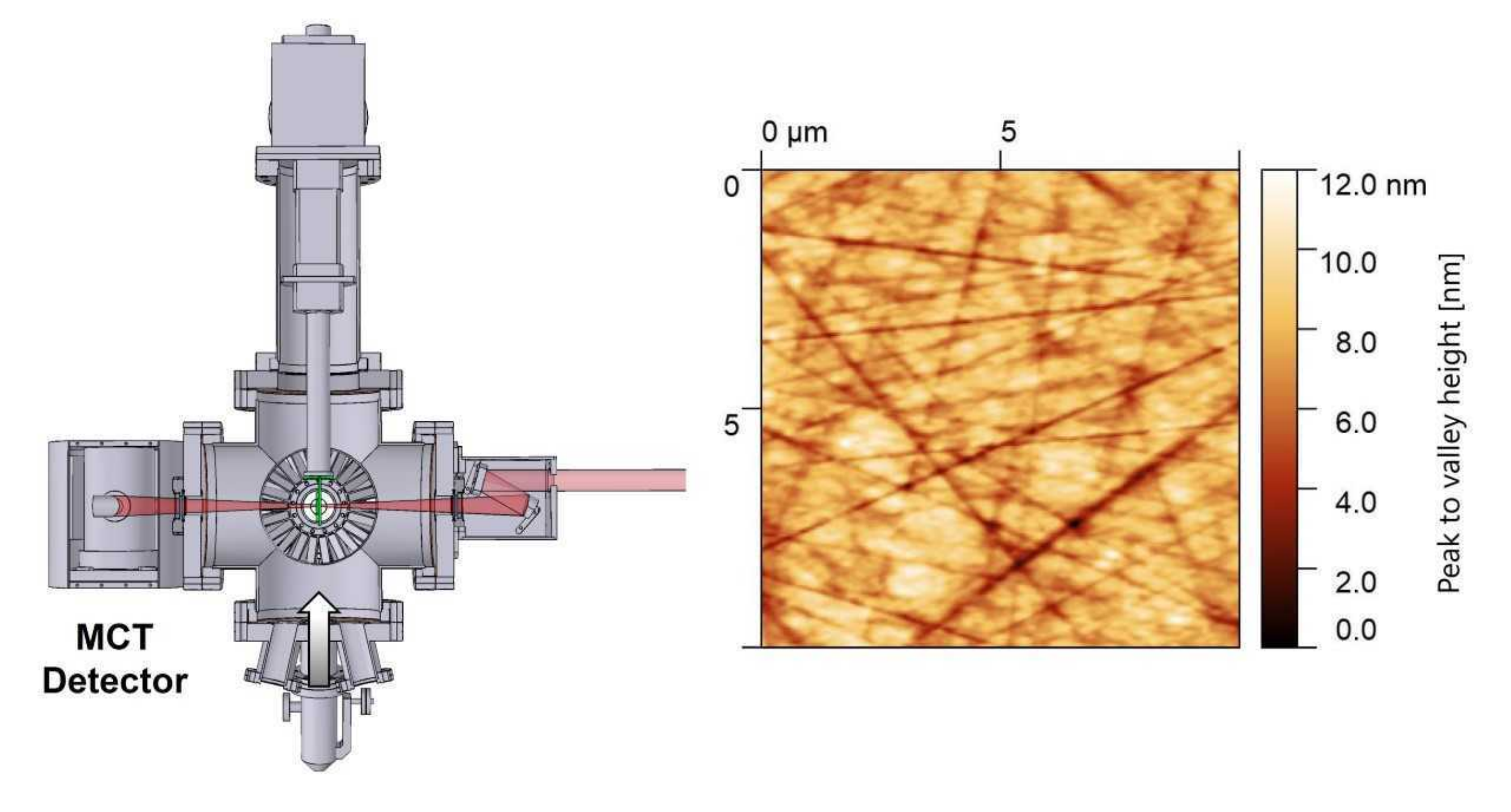}
\caption{Left part: a schematic view of the ISEAge UHV setup. The IR beam path is shown with red. The T-shaped optical holder with the Ge window is shown with green. The direction of molecular species injection is shown with arrow. Cad models are taken from Kurt J. Lesker catalogue$^a$. The radiation shield is omitted for clarity. Right part: a typical image of Ge window surface obtained with the scanning-probe microscope for the 10x10 $\mu$m  area. The colour scheme indicates the registered peak to valley height from the lowest point (dark brown) to the highest point (light yellow) in nm.}
$^a$ \url{http://www.lesker.com/vacuum-flanges-components.cfm/}\\\label{fig:fig1}
\end{figure*}

All experiments are performed using Ice Spectroscopy Experimental Aggregate (ISEAge), a custom designed ultra-high vacuum (UHV) setup, see Figure \ref{fig:fig1}. A symmetrical stainless steel 6-way cross is utilized as a main UHV chamber. Ices are grown with a nanometer precision on both sides of a 10~\texttimes~10~mm Ge window located at the center of the main chamber. A “T-shape” gold-coated optical holder is used to attach Ge window to the cold finger of a vertically placed Janis RDK-205J cryostat. A horizontally attached VCMO KVT-400 turbomolecular pump in combination with the oil-free Kashiyama NeoDry 15E multi-stage roots forevacuum pump is used to obtain the pressure as low as 2~\texttimes~10$^{-10}$~mbar prior to the deposition. Molecular species are introduced through the two symmetrically placed all-metal leak valves located in the same plane as the Ge window at the bottom of the main chamber. Custom made gold-coated radiation shield placed along the paths of the injected molecular species and the utilization of symmetrical geometry of the setup secure equal deposition of the species on both sides of the Ge window through the so-called “background deposition” \citep{Rachid_et_al._2021,Rachid_et_al._2022}.

LakeShore DT‐670B‐SD standard curve Silicon diode is used to monitor the temperature of the optical holder that is in the thermal equilibrium with the Ge window. The temperature of the optical holder is regulated in the range between 6.5 and 305 K by Lakeshore 335 controller through the application of the resistive (50 $\Omega$) heating. Relatively good thermal conductivity properties of Ge at low temperatures \citep{Carruthers,levinshtein1999handbook} in combination with the compact dimensions of the Ge window are utilized to minimize the temperature gradients over the Ge window radius. The 6 mm diameter circle area of the window is available for the deposition of the ice. 

Thermo Scientific Nicolet iS50 FTIR spectrometer is used to acquire transmittance IR spectra of the grown ices in the range between 4000 and 630~cm$^{-1}$ (2.5 and 15.9~$\mu$m) with 1~cm$^{-1}$ resolution. The focused IR beam is introduced into the main chamber through a ZnSe viewport with the focal point centered on the Ge window. The outgoing IR beam is then passed through another ZnSe viewport and focused on the L-N$_2$ cooled HgCdTe detector. FTIR spectrometer and the whole path of the IR beam outside of the UHV main chamber is flushed with pure N$_2$ generated by Himelectronika GA-600 N$_2$ generator on a 24/7 basis.

The background deposition rates of the injected molecular species are regulated by controlling their corresponding partial pressures in the main chamber of the setup by means of the Stanford Research Systems RGA200 quadrupole mass spectrometer (QMS), see for example \cite{Slavicinska_et_al._(2023)} for the description of the similar procedure. The QMS is situated in the front part of the setup opposite to the turbomolecular pump. Application of the QMS ion source with the adjustable electron energy regulated in the range between 25 and 105~eV allows for the deconvolution of the species with overlapping m/z ion signals, such as HCN and C$_6$H$_6$ or C$_6$H$_5$NH$_2$.

Two isolatable all-metal dosing lines are used to contain vapours of the injected species for the whole time of the experiments. The dosing lines are supplied with the shared gas mixing line and a gas buffer. All parts of the dosing line system are prepumped with the VCMO KVT-50 turbopump and an independent oil-free Kashiyama NeoDry 15E forevacuum pump to the pressure \textless10$^{-4}$~mbar.  Application of the ConFlat flanges and all-metal UHV valves for the design of the dosing line system allows for the stable depositions of species with the vapour pressures of the order of 10$^{-1}$~mbar, such as C$_6$H$_5$NH$_2$. Purity of the injected species is continuously monitored by using QMS.
The Ge window surface properties are characterized using ex situ NTEGRA MFM scanning-probe microscope, see right panel of Figure \ref{fig:fig1} for the example of obtained data. The maximum peak to valley height parameter of 12~nm is measured on 10~\texttimes~10~$\mu$m surface sample. This means that the surface can be considered optically flat for the wavelength range used in present study. Moreover, a full surface coverage can be assumed for the ice thicknesses above 10~nm. Up to our knowledge this is the first time when the surface properties are characterized prior to the acquisition of interstellar ice analogue mid-infrared spectroscopic data. 

The compounds used are as follows: CO (99.9999~\%, Ugra-PGS), CO$_2$ (99.9999~\%, Ugra-PGS), NH$_3$ (99.9~\%, UralKrioGaz), deionized H$_2$O, CH$_3$OH ($\geqslant$99.8~\%, Vekton), C$_6$H$_6$ ($\geqslant$99.8~\%, Vekton), C$_5$H$_5$N ($\geqslant$99.8~\%, Vekton), C$_6$H$_5$NH$_2$ ($\geqslant$99.8~\%, Component-reactiv), K$_4$[Fe(CN)$_6$]$\times$3H$_2$O ($\geqslant$99.8~\%, Vekton), C$_{17}$H$_{35}$COOH (stearic acid) ($\geqslant$99.8~\%, Vekton). HCN is obtained “in situ” in the dosing line of the UHV setup following the well-established methodology described by \cite{Gerakines_Moore_&_Hudson_(2004)} and \cite{10.1093/mnras/sts272}. Gaseous CO and CO$_2$ are introduced directly into the dosing lines from the commercially acquired gas bottles. CH$_3$OH, C$_6$H$_6$, C$_5$H$_5$N and C$_6$H$_5$NH$_2$ vapours are obtained from their corresponding liquid samples. Each of the used liquid samples are preliminary degassed using three consequent freeze-pump cycles.

\section{Methods} \label{sec:methods}

All key experiments are performed in the following way. At first pure HCN is introduced into the main UHV chamber with the fixed deposition rate simultaneously controlled by the QMS in the gas phase and by the FTIR at the substrate surface. The deposition is performed for the period of 60 minutes at 10~K substrate temperature with the controlled rate equal to 2.7~\texttimes~10$^{12}$~cm$^{-2}$~s$^{-1}$. Upon acquisition of 1~\texttimes~10$^{16}$~cm$^{-2}$ HCN column density, second gas (or NH$_3$:H$_2$O gas mixture) is introduced into the main chamber through the second independent leak valve simultaneously with the continuous deposition of HCN molecules. The second ice component deposition rate is controlled with the QMS, while the co-deposition ratio is set to 1:8 (1:3:8 for the case of HCN:NH$_3$:H$_2$O ice mixture). With the application of 1:8 ratio, a single HCN molecule is surrounded on average by a cube of 8 lattice molecules. This aims to provide a separation between the HCN molecules in the ice. On the other hand, the use of 1:8 ratio allows to keep the partial pressure of the second deposited ice component within the measurement range of the ion source of the QMS for the whole time of co-deposition. The co-deposition is continued for another 60 minutes. This technique allows for the exact reproduction of HCN deposition rate and the obtained HCN column density in all of the performed experiments on a day-to-day basis. By estimating the IR absorption band areas for the equal amounts of HCN molecules embedded in various molecular environments, the relative absorption band strength values of HCN can be directly evaluated. It should be noted that only the last 30 minutes of co-deposition are used for the analysis to avoid any impact of the phase interference between pure and mixed HCN ice layers. 
The selected experiments are repeated using temperature programmed desorption technique. In this case the ice mixture is obtained with the very same technique, but without the pre-growth of the pure HCN ice layer. Upon acquisition of this mixed ice and comparison of the result of co-deposition with the result of the original key experiment, the ice mixture is warmed up with 5 K per minute rate. 
The IR spectra are continuously collected and recorded every 45 seconds (averaging of 32 scans) for the whole time of (co)deposition and ice warm up. Separate calibration curve between the m/z ion signal of the QMS and the column density of the grown ice is obtained for each of the molecules used in the experiments, see, for example \cite{Slavicinska_et_al._(2023)}. The band strength values used for the acquisition of the calibration curves are presented in Table \ref{tab:table1}. The preference was given to the band strength values obtained at the similar substrate temperature. Alternatively, the values which are reproducible by the literature were chosen.

\begin{deluxetable*}{cccc}
\tablenum{1}
\tablecaption{The list of used absorption band strength values\label{tab:table1}}
\tablewidth{0pt}
\tablehead{
\colhead{Molecule} & \colhead{Vibration Mode} &
\colhead{Band Strength} & \colhead{References} \\
\colhead{} & \colhead{} &
\colhead{(cm)} & \colhead{}
}
\startdata
HCN & $\nu_3$ & 1.029\texttimes10$^{-17}$ & \cite{Gerakines_Yarnal_and_Hudson_(2022)} \\
H$_2$O & $\nu_3$ &  2.2\texttimes10$^{-16}$ & \cite{Bouilloud2015} \\
NH$_3$ & $\nu_2$ & 1.63\texttimes10$^{-17}$ &  \cite{Bouilloud2015} \\
CO & $\nu_1$ & 1.12\texttimes10$^{-17}$ & \cite{Bouilloud2015} \\
CO$_2$ & $\nu_3$  & 1.1\texttimes10$^{-16}$ & \cite{Bouilloud2015} \\
CH$_3$OH & $\nu_8$ & 1.78\texttimes10$^{-17}$ & \cite{Bouilloud2015} \\
C$_6$H$_6$ & $\nu_{13}$ ($\nu_{19}$) & 4.80\texttimes10$^{-18}$ & \cite{HUDSON2022114899} \\
C$_5$H$_5$N & $\nu_{22}$ & 7.47\texttimes10$^{-18}$ & \cite{HUDSON2022114899}  \\
C$_6$H$_5$NH$_2$& $\nu_{25}$  & 1.1\texttimes10$^{-17}$ & This work$^a$ \\
\enddata
\tablecomments{$^a$ Evaluated under assumption that the ratio between C$_6$H$_5$NH$_2$ and C$_6$H$_6$ deposition rates is proportional to the ratio between the molecular velocities of these species.}
\end{deluxetable*}

\section{Results and Discussion} \label{sec:discussion}

\begin{figure}[ht!]
\includegraphics[scale=0.5]{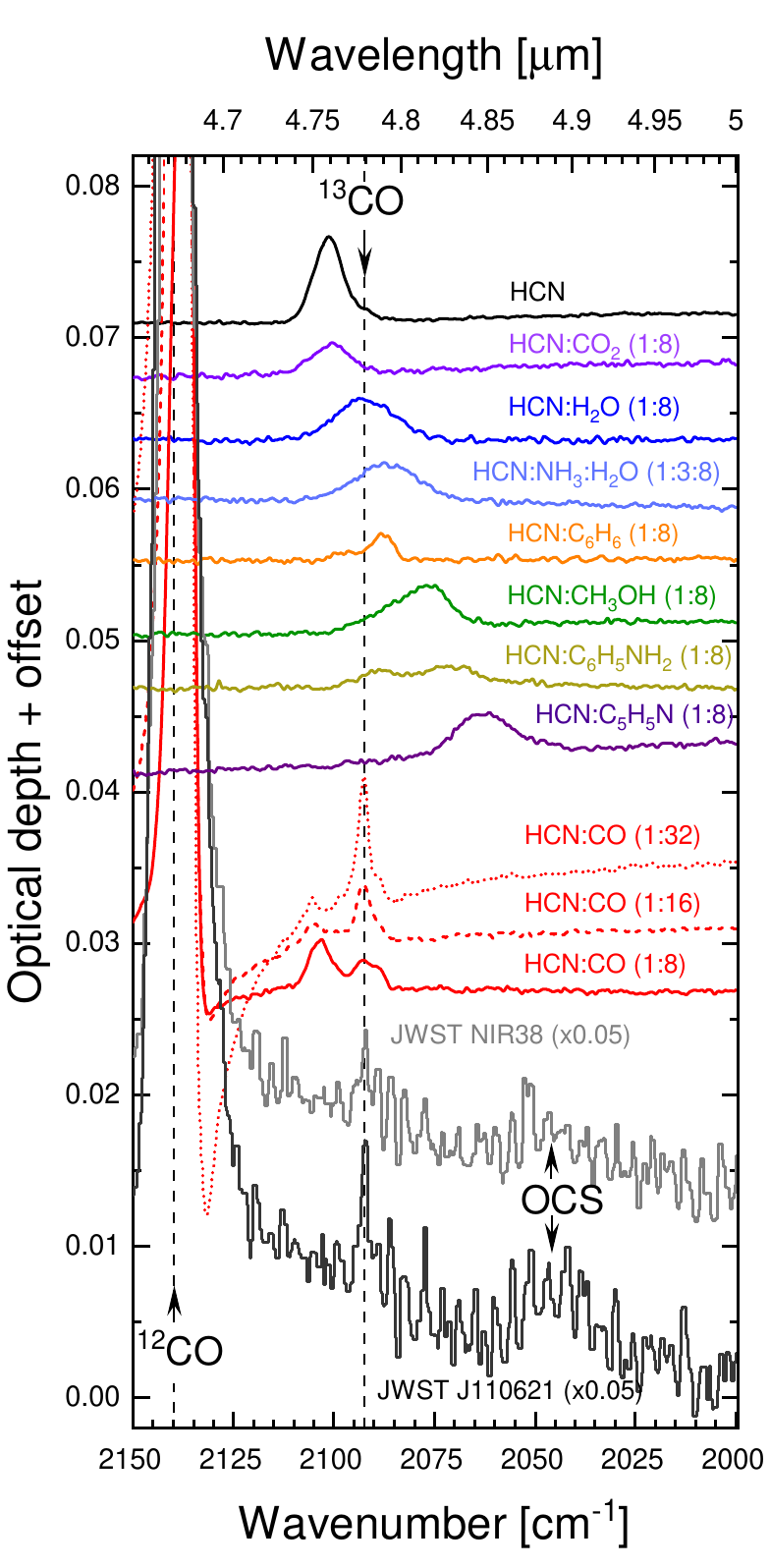}
\centering
\caption{Top part: the position and profile of ($\nu_3$) absorption feature of HCN molecules embedded in various astrochemically relevant molecular environments, including most abundant molecular constituents of interstellar ices. All spectra are obtained using ISEAge UHV setup at 10~K substrate temperature. The spectra presented with solid lines are obtained in the experiments using identical HCN deposition rate and the total time of co-deposition. The column density of HCN molecules is equal to 5~\texttimes~10$^{15}$~cm$^{-2}$ for each of the experiments. Bottom part: the JWST (NIRSpec) IR spectra of pristine cloud ices towards two background stars, NIR38 and J110621.63, using JWST (NIRSpec) presented for comparison \citep{McClure_et_al._2023}. Spectra are presented with an offset. The JWST spectra are multiplied by a factor of 0.05 for clarity. Consequently, the column density of HCN in the laboratory experiments should be treated as 1~\texttimes~10$^{17}$~cm$^{-2}$ while comparing with the data from JWST. This corresponds to the abundance of $\thicksim$1~\% with respect to the H$_2$O ice \citep{McClure_et_al._2023}. The peak frequencies of HCN ($\nu_3$) absorption feature in various molecular environments are reported in the Table \ref{tab:tableA1} of Appendix A.
\label{fig:fig2}}
\end{figure}

The full range IR spectra acquired in the present study are available at the following
URL: \url{https://doi.org/10.5281/zenodo.10730932}. In Figure \ref{fig:fig2} a comparison between the ($\nu$$_3$) CN-stretching mode absorption features of HCN molecules embedded in various molecular environments is presented. For the clarity purposes the JWST NIRSpec FS spectra of pristine cloud ices towards two background stars, NIR38 and SSTSL2J110621.63-772354.1 (J110621) from \cite{McClure_et_al._2023} are presented. Experimental data show that both the position and the profile of the ($\nu$$_3$) absorption feature depends strongly on the molecular environment. The red shift as much as 36.5~cm$^{-1}$ is observed in comparison to the peak position for the pure HCN ice. The largest shift is detected for the HCN molecules embedded into CH$_3$OH,  C$_6$H$_5$NH$_2$ and C$_5$H$_5$N ices. 

For the HCN molecules embedded into C$_6$H$_6$ and C$_6$H$_5$NH$_2$ ices a double peak structure is observed. The double peak structure of the HCN absorption features in C$_6$H$_6$ environment is presumably caused by the partial segregation of HCN molecules in the ice comprised of large flat C$_6$H$_6$ molecules. The absorption feature centered at 2098~cm$^{-1}$ is tentatively assigned to the clusters of HCN molecules, while the feature at 2088~cm$^{-1}$ is tentatively assigned to the HCN molecules isolated in C$_6$H$_6$. This interpretation is based on the observed decrease of 2088~cm$^{-1}$ absorption feature and the simultaneous increase of the 2098~cm$^{-1}$ absorption feature following the partial segregation of the ice upon warm up, see Appendix B for more details. It should be noted that C$_6$H$_6$ and C$_6$H$_5$NH$_2$ environments aim to somehow mimic the interaction of HCN molecule with the aromatic rings of the PAHs which supposedly comprise considerable part of the carbonaceous material of the grains \citep{Tielens_2008}.

The best match between the peak position and profile of the 4.78~$\mu$m (2090~cm$^{-1}$) feature, detected on the ice spectra towards NIR38 and J110621, and the laboratory spectroscopic data are observed for the absorption feature of naturally abundant $^{13}$CO isotopologue molecules contained in HCN:CO ice mixtures. This is in line with the current assignment of 4.78~$\mu$m absorption feature to the $^{13}$CO ice by \cite{McClure_et_al._2023} and is further constrained by the unambiguous observations of $^{12}$CO ice towards NIR38 and J110621. However, based on our laboratory results, attribution of the part of 4.78~$\mu$m feature to HCN molecules cannot be completely excluded. Indeed, a comparison between the laboratory data and the ice spectra towards NIR38 and J110621, shows that a part of the 4.78~$\mu$m feature can be attributed to the absorption of HCN molecules embedded in H$_2$O environment. The presence of naturally abundant $^{13}$CO isotope in HCN:CO ice mixtures complicates the assignment of HCN ($\nu$3) absorption feature in the experiments. Currently, the 2104 cm$^{-1}$ absorption feature observed in the spectra of HCN:CO ice mixtures can be tentatively assigned to the various clusters of HCN molecules embedded in CO ice. This is consistent with the gradual decrease of 2104 cm$^{-1}$ absorption feature area upon the dilution with CO molecules. The experiments with HCN:$^{12}$CO ice mixtures are required to draw some solid conclusions on the possible contribution of HCN:CO absorption into the 4.78~$\mu$m feature observed with JWST. The JWST data with a better signal-to-noise ratio as well as the accurate knowledge of the $^{13}$CO/$^{12}$CO isotopic ratio are required to derive the precise upper limits on the amount of HCN ice embedded in various molecular environments towards NIR38 and J110621. Currently, the maximum relative abundance of about 1\% with respect to the H$_2$O reported by \cite{McClure_et_al._2023} is a reasonable upper limit on the amount of HCN in the ices towards NIR38 and J110621.

\begin{figure}[ht!]
\includegraphics[scale=0.5]{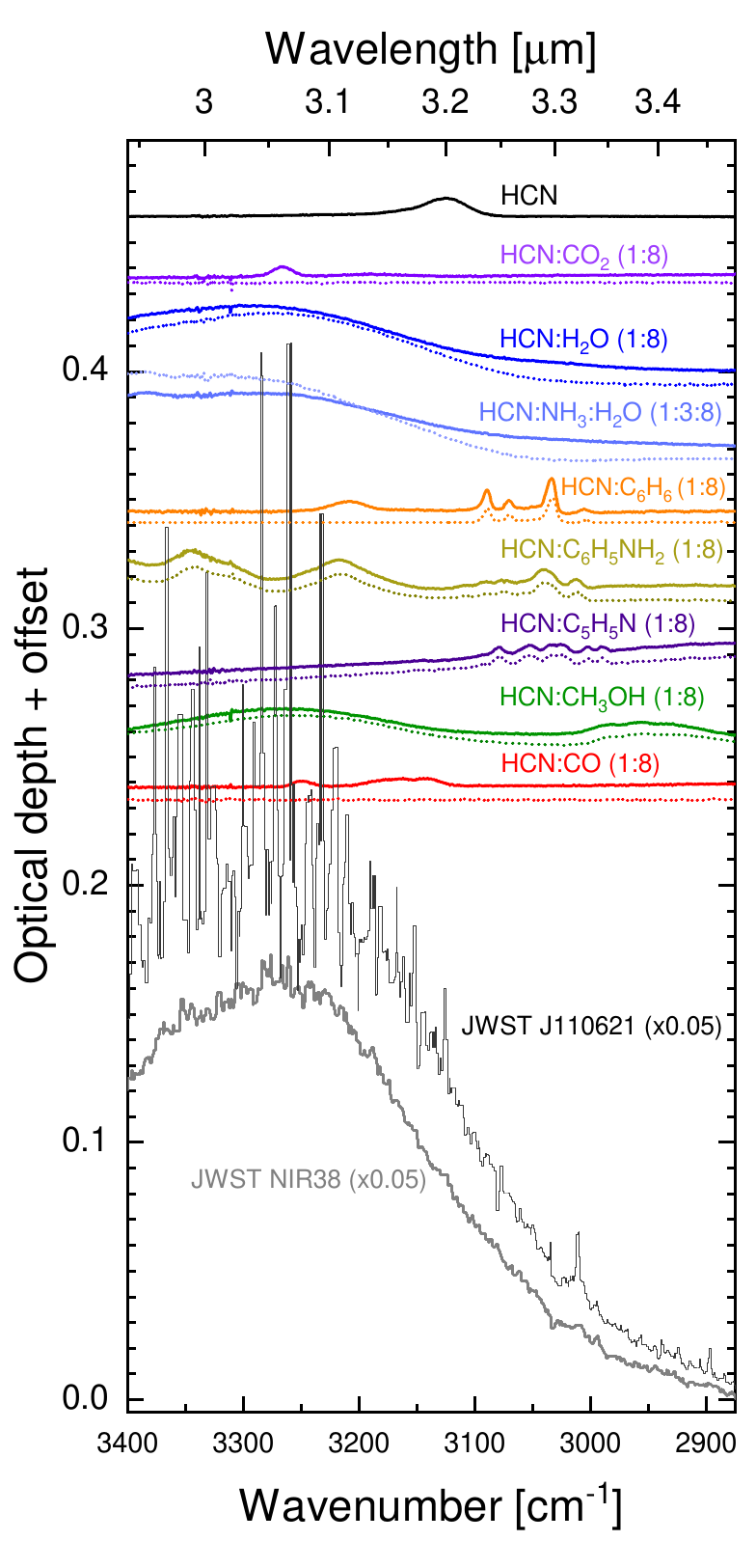}
\centering
\caption{Similar to Figure \ref{fig:fig2}, a comparison between the position and profile of ($\nu_1$) absorption feature of HCN molecules. The dotted spectra are obtained for the pure molecular environment without addition of HCN molecules and are presented for the reference.  
\label{fig:fig3}}
\end{figure}

An analogical comparison for the ($\nu_1$) CH-stretching mode absorption features of HCN molecules in various molecular environments is presented in Figure \ref{fig:fig3}. The position of ($\nu_1$) absorption feature demonstrates even larger shift depending on the environment. This is not surprising, considering the weaker energy of the single CH bond in comparison to the triple CN bond. The shift as high as 140~cm$^{-1}$ can be observed for the HCN:CO$_2$ ice. Under our experimental conditions CH-stretching mode absorption feature of HCN cannot be resolved in the experiments with H$_2$O, H$_2$O:NH$_3$, CH$_3$OH and C$_6$H$_5$NH$_2$ environments. This limits the use of $\nu_1$ absorption feature for identification of HCN in the interstellar ices. Despite the higher band strength value, the strong overlap between the ($\nu_1$) absorption features of HCN in various environments with the absorption features caused by the OH- and NH-stretching modes of the abundant H$_2$O, NH$_3$, NH$_4^+$ and CH$_3$OH ices prohibits HCN identification towards NIR38 and J110621. 

\begin{figure}[ht!]
\includegraphics[scale=0.32]{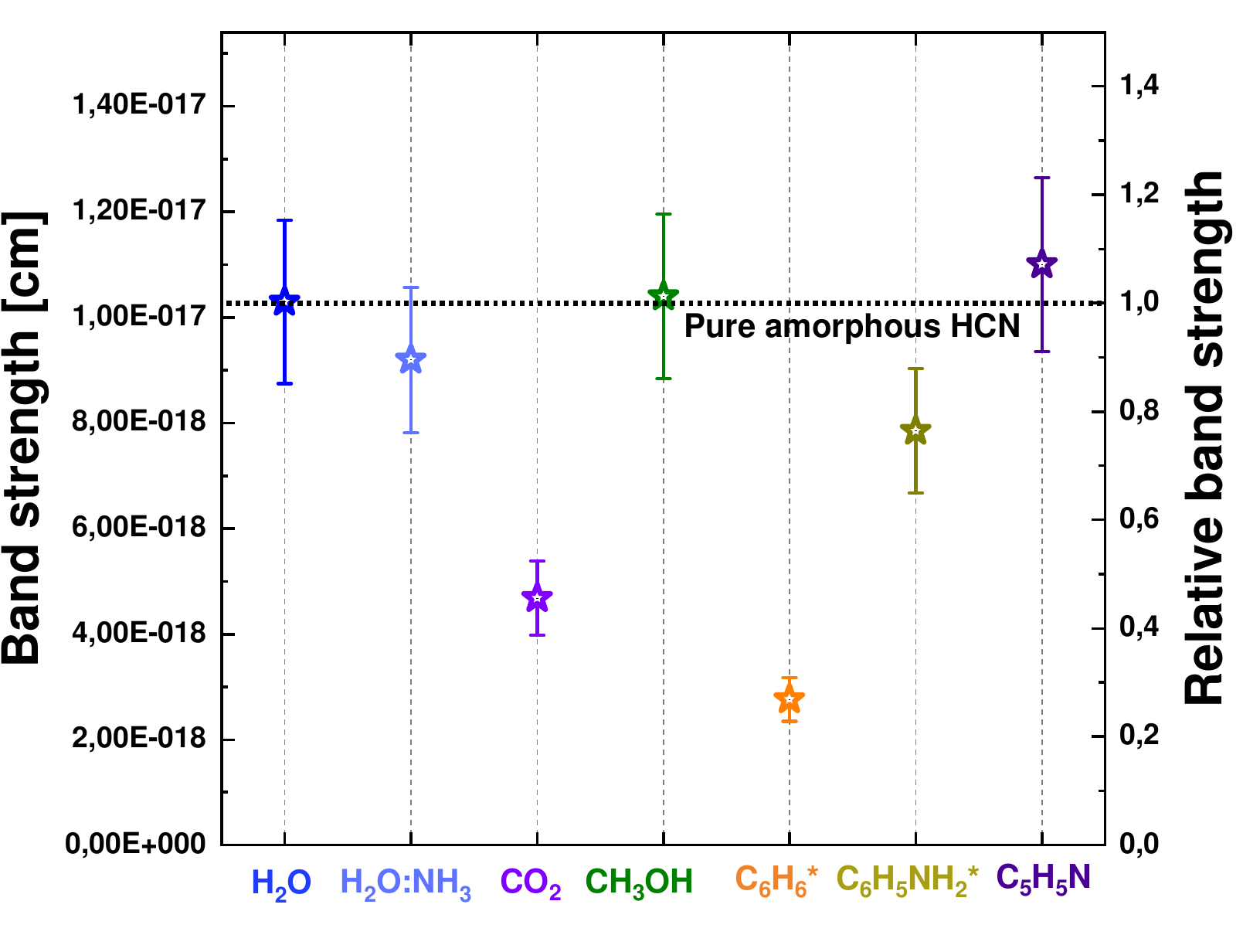}
\centering
\caption{The relative HCN ($\nu_3$) band strength values in different molecular environments. The value for pure HCN ice is adopted from \cite{Gerakines_Yarnal_and_Hudson_(2022)}. The tentative values are marked with asterisks. The corresponding band strength values obtained for HCN ($\nu_3$) absorption feature in various molecular environments are reported in the Table \ref{tab:tableA1} of Appendix A.
\label{fig:fig4}}
\end{figure}

All key experiments presented in Figure \ref{fig:fig2} and \ref{fig:fig3} are performed using the identical deposition rate and deposition time of HCN molecules. This allows for the direct comparison of the relative band strength values of HCN in different environments. The evaluated data are presented in Figure \ref{fig:fig4}. Our results confirm previous findings reported by \cite{Gerakines_Yarnal_and_Hudson_(2022)}. Near equal HCN ($\nu_3$) band strength values are observed for the pure HCN and the mixed HCN:H$_2$O ices.

It should be noted that the estimation of the HCN band strength in the mixed NH$_3$:H$_2$O ice should be done with care. A conversion of HCN into the CN$^{-}$ anion can partially occur following the acid-base reaction with NH$_3$ well investigated by \cite{10.1093/mnras/sts272}. Under the experimental conditions utilized in the present work, with the HCN:NH$_3$:H$_2$O mixing ratio equal to 1:3:8, the onset of HCN conversion into CN$^{-}$ is observed upon the warm up of the mixed ice above 40~K. This chemical conversion is observed through the shift of the ($\nu_3$) HCN absorption feature at 2087~cm$^{-1}$ to the 2084~cm$^{-1}$ absorption feature of CN$^{-}$ anion and simultaneous appearance of the broad absorption feature of NH$_4^+$ centered at 1487~cm$^{-1}$, see \cite{10.1093/mnras/sts272}. See also Appendix C for more details.

Similarly, the band strength estimations for HCN in C$_6$H$_6$ and C$_6$H$_5$NH$_2$ environments are complicated by the double-peak structure of HCN ($\nu_3$) absorption feature, see Appendix A for more details. The exact ratio between the two contributions into the total area of HCN ($\nu_3$) absorption feature can vary depending on the applied physical conditions. The total area of ($\nu_3$) absorption feature of HCN molecules is used to estimate the band strength values of HCN isolated in the solid C$_6$H$_6$ and C$_6$H$_5$NH$_2$. Therefore, such estimations can only be considered as tentative. 

\section{Conclusion} \label{sec:conclusion}

In order to facilitate the search for interstellar HCN ice using JWST the series of systematic laboratory measurements is performed to acquire mid-infrared spectra of HCN molecules embedded in various astronomically relevant environments. These environments include the main constituents of icy mantles of interstellar grains, i.e., H$_2$O, NH$_3$, CO, CO$_2$ and CH$_3$OH, or aromatic environments comprised of solid C$_6$H$_6$, C$_5$H$_5$N and C$_6$H$_5$NH$_2$. The latter aims to somehow resemble the carbonaceous material of the interstellar grains. 
The comparison between the acquired laboratory data and the JWST (NIRSpec) data towards quiescent molecular clouds available from \cite{McClure_et_al._2023} shows that the partial assignment of HCN absorption to 4.78~$\mu$m band cannot be excluded along with the current attribution to $^{13}$CO ice. Overall, the NIRSpec data towards brighter sources with the higher total ice column densities are highly required to draw solid conclusions on the possible presence of HCN in the interstellar ices. Such data can come out with the realization of the GO program which acquired a JWST NIRSpec spectroscopic data in the range between $\thicksim$2.87~$\mu$m and $\thicksim$5.27~$\mu$m for a number of protostars, see \cite{nazari2024hunt}.
With the goal to facilitate further observations and support the search of interstellar HCN ice the acquired full-range laboratory spectroscopic data are made publicly available at the following URL: \url{https://doi.org/10.5281/zenodo.10730932}

\begin{acknowledgments}
This research work is funded by the Russian Science Foundation via the agreement 23-12-00315, and by the Russian Ministry of Science and Higher Education via the State Assignment Contract FEUZ-2020-0038 (Appendix B). We express our deep gratitude to Sergio Ioppolo, Ko-Ju Chuang and Martijn Witlox for sharing their valuable experience in construction and operation of UHV apparatuses. We are grateful to Dmitrii Obydennov for sharing his deep expertise in organic chemsitry with us. We also would like to thank Alexey Mozhegorov and Andrey Ostrovsky  for the stimulating discussions.
\end{acknowledgments}

\newpage
\appendix

\section{Appendix}
\begin{deluxetable}{ccc}[h]
\tablenum{A1}
\tablecaption{The list of observed peak frequencies and band strengths of HCN ($\nu_3$) absorption feature in various molecular environments. The estimated errors of band strengths are $\leq$ 15 \%
\label{tab:tableA1}}
\tablewidth{0pt}
\tablehead{
\colhead{Ice mixture} & \colhead{Peak frequency} &
\colhead{Band Strength} \\
\colhead{} &
\colhead{(cm$^-$$^1$)} & \colhead{cm}
}
\startdata
HCN:H$_2$O & 2092.5 & 1.0\texttimes10$^{-17}$ \\
HCN:NH$_3$:H$_2$O & 2085.6 & 9.2\texttimes10$^{-18}$ \\
HCN:CO$_2$ & 2100.3 & 4.7\texttimes10$^{-18}$ \\
HCN:CH$_3$OH & 2076.2 & 1.0\texttimes10$^{-17}$ \\
HCN:C$_6$H$_6$ & 2095.7, 2087.9 & 2.8\texttimes10$^{-18a}$ \\
HCN:C$_6$H$_5$NH$_2$ & 2088.0, 2071.8 & 7.8\texttimes10$^{-17a}$ \\
HCN:C$_5$H$_5$N & 2063.5 & 1.1\texttimes10$^{-17}$ \\
\enddata
\tablecomments{$^a$ The combined area of both absorption features is used for evaluation}
\end{deluxetable}

\section{Appendix}
\begin{figure}[ht!]
\includegraphics[scale=0.5]{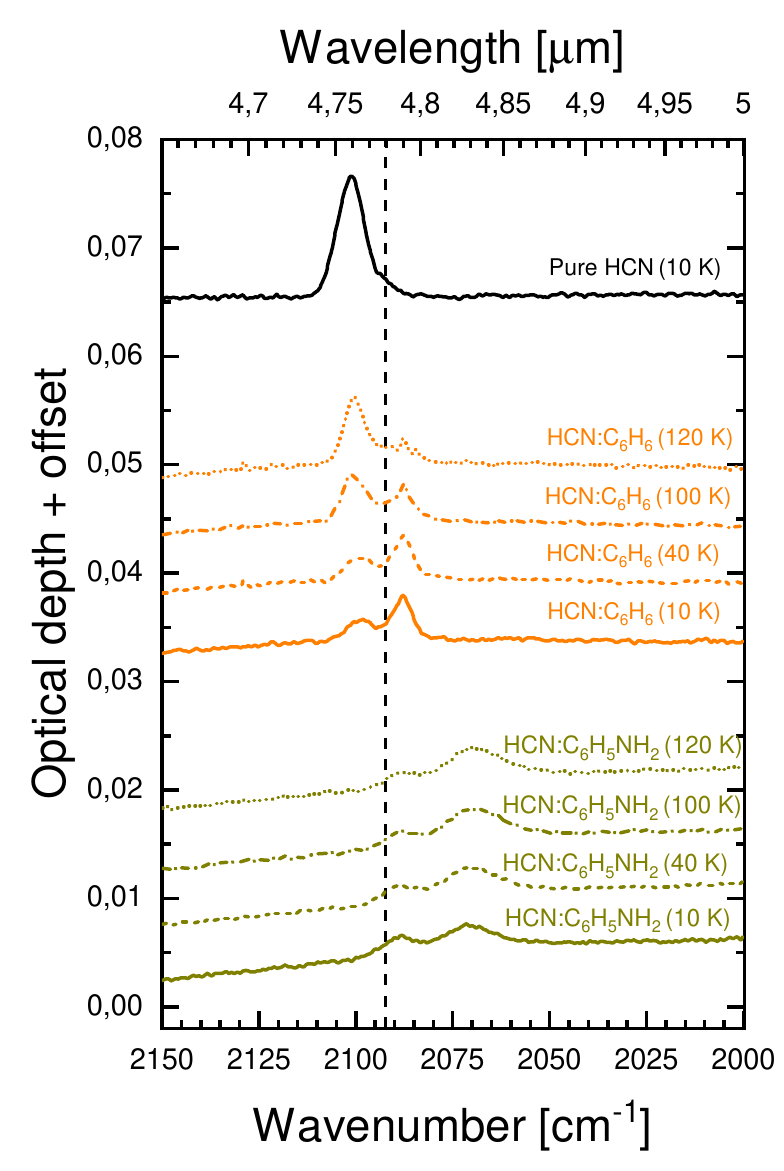}
\centering
\caption{The evolution of the position and profile of ($\nu_3$) absorption feature of HCN molecules embedded in C$_6$H$_6$ and C$_6$H$_5$NH$_2$ ices with the ratio of 1 to 8 upon the ice warm up. The initial column density of HCN molecules is equal to $\thicksim$1~\texttimes~10$^{16}$~cm$^{-2}$ for each of the experiments. Spectra are presented with an offset. The spectrum of pure HCN ice is presented for a comparison. The dashed vertical line indicates the position of 4.78~$\mu$m band identified on the JWST (NIRSpec) spectra and assigned to $^{13}$CO, see \cite{McClure_et_al._2023}. 
\label{fig:fig5}}
\end{figure}

The 2150–2000~cm$^{-1}$ region of the IR spectra acquired during the temperature programmed warm up of HCN:C$_6$H$_6$ (1:8) and HCN:C$_6$H$_5$NH$_2$ (1:8) ices is presented in the Figure \ref{fig:fig5}. The clear evolution of the HCN ($\nu_3$) double peak absorption feature can be observed for HCN:C$_6$H$_6$ ice mixture with the temperature increase. The shape and the position of the rising left component of this double peak absorption feature closely resembles the shape and position of the ($\nu_3$) absorption feature of the pure HCN ice. This evolution of the absorption feature profile can be explained by the partial segregation of the ice upon the temperature increase. Indeed, formation of HCN clusters in the ice should give a rise to the absorption feature of HCN molecules surrounded by other HCN molecules, while the area of the right component of the double peak, attributed to the absorption of HCN molecules encaged by C$_6$H$_6$ molecules, should decline. This segregation can be expected given the large size of C$_6$H$_6$ molecules with respect to HCN and the lack of strong interactions between C$_6$H$_6$ molecules and HCN molecules, e.g., hydrogen bonds. On the contrary to C$_6$H$_6$ molecules, the molecules of C$_6$H$_5$NH$_2$ can form hydrogen bonds with HCN. This may explain the lack of segregation and the associated change in the HCN ($\nu_3$) absorption feature profile upon the HCN:C$_6$H$_5$NH$_2$ (1:8) ice mixture warm up, see Figure \ref{fig:fig5}. The presence of the two distinct components on the absorption feature of HCN molecules embedded in solid C$_6$H$_5$NH$_2$ is likely caused by the two different orientations of HCN molecules with respect to C$_6$H$_5$NH$_2$ in this mixed ice.

\section{Apendix}

\begin{figure}[ht!]
\includegraphics[scale=0.38]{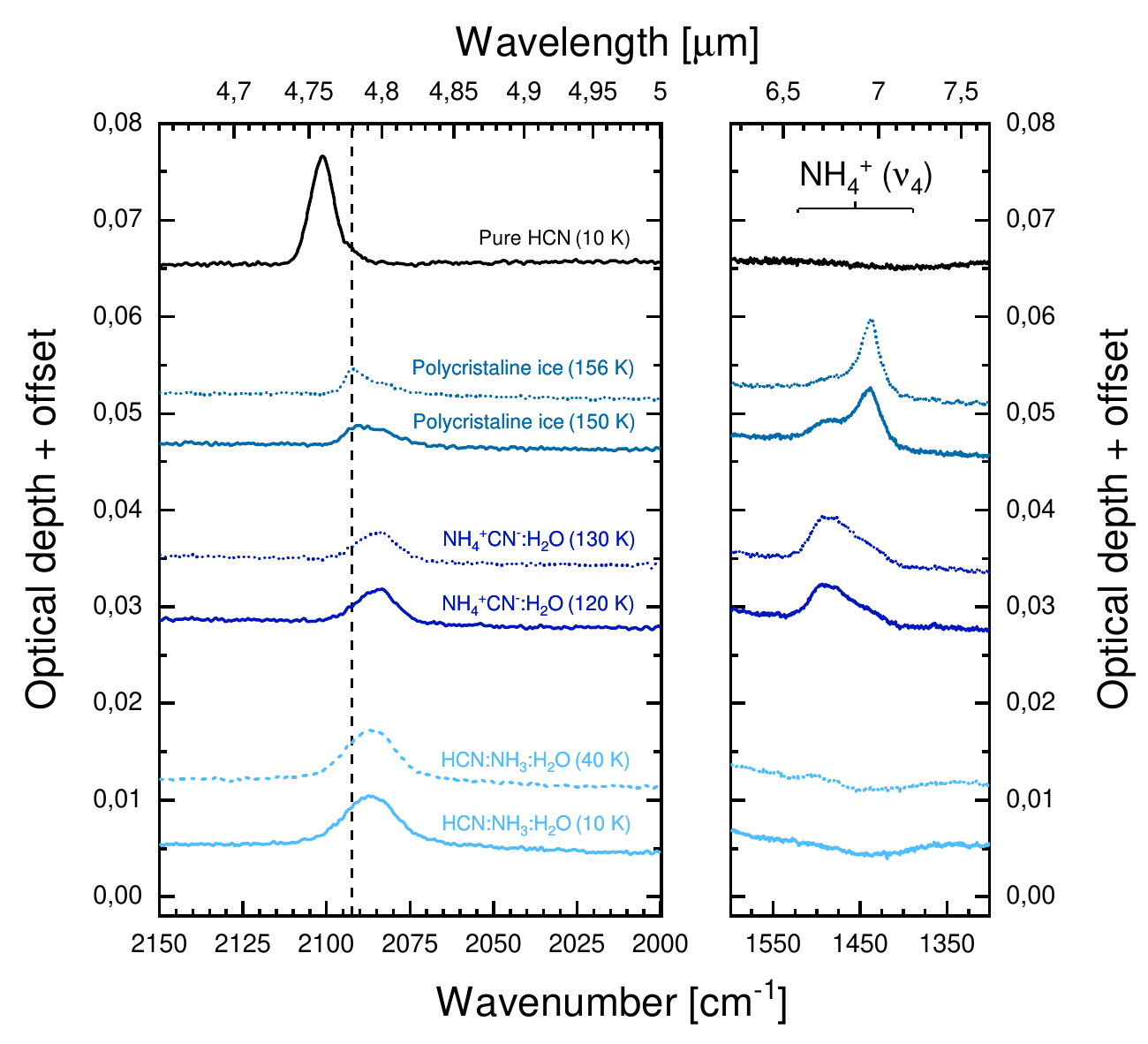}
\centering
\caption{The fragments of IR spectra obtained upon the mixed HCN:NH$_3$:H$_2$O (1:3:8) ice warm up. Left: evolution of the position and profile of 2087~cm$^{-1}$ absorption feature. Right: appearance and evolution of the NH$_4^+$ absorption feature. The initial column density of HCN molecules is equal to $\thicksim$1~\texttimes~10$^{16}$~cm$^{-2}$ . Spectra are presented with an offset. The spectrum of pure HCN ice is presented for a comparison. The dashed vertical line indicates the position of 4.78~$\mu$m band identified on the JWST (NIRSpec) spectra and assigned to $^{13}$CO, see \cite{McClure_et_al._2023}. 
\label{fig:fig6}}
\end{figure}

The fragments of IR spectra acquired during the temperature programmed warm up of the HCN:NH$_3$:H$_2$O (1:3:8) ice are presented in Figure \ref{fig:fig6}. Formation of NH$_4^+$CN$^-$ following the reaction NH$_3$ + HCN → NH$_4^+$CN$^-$ is observed upon the temperature processing of the ice in agreement with the results reported by \cite{10.1093/mnras/sts272}. This can be observed through the appearance of the broad NH$_4^+$ absorption feature centered around 1487~cm$^{-1}$ at the temperature above 40~K, and the evolution of the shape and position of the absorption feature at 2087~cm$^{-1}$. The latter marks the conversion of CN-stretching mode absorption feature of HCN into the absorption feature of CN$^-$. Following the crystallization of the water ice at the temperatures above 140~K \citep{Hudgins1993,Lamberts2015} the consequent changes in the position and profile of CN$^-$ and NH$_4^+$ absorption features are observed, see \cite{10.1093/mnras/sts272}.

As NH$_3$ molecules are presented in overabundance with respect to HCN molecules, a close to total conversion of HCN into NH$_4^+$CN$^-$ can be expected upon the warm up of the mixed ice. Thus, a value for the relative band strength of CN$^-$ with respect to the band strength of HCN molecules ($\nu_3$) can be derived from the change in the total area of the 2087~cm$^{-1}$ absorption feature. The value of 2.1 is obtained for the ratio between the peak areas at 10 and 130 K. Adoption of HCN band strength values presented in \cite{Gerakines_Moore_&_Hudson_(2004)} and in Figure \ref{fig:fig4} of this study results in the CN$^-$ absorption band strength equal to 4.3~\texttimes~10$^{-18}$~cm. The estimate reported by \cite{10.1093/mnras/sts272} for the CN$^-$ absorption band strength is equal to 1.8~$\pm$~1.5~\texttimes~10$^{-17}$~cm. The value of 4.3~\texttimes~10$^{-17}$ is about 4 times lower than the estimate evaluated by \cite{10.1093/mnras/sts272}, however it lies within the high reported uncertainty of their measurements.

It should be noted that the position of CN$^-$ absorption feature of NH$_4^+$CN$^-$ in the polycrystalline ice at 150~K closely resembles the position of 4.78~$\mu$m band identified on the JWST (NIRSpec) spectra and assigned to $^{13}$CO, see \cite{McClure_et_al._2023}. However, the spectra presented by \cite{McClure_et_al._2023} do not show signs of temperature processing and crystallization of H$_2$O ice. Thus, the assignment of 4.78~$\mu$m band to NH$_4^+$CN$^-$ embedded in either amorphous or polycrystalline ice is dubious.  On the other hand, the column density of NH$_4^+$ ions reported by \cite{McClure_et_al._2023} is considerably higher than the column density of OCN$^-$ anions reported in their study, indicating the lack of negatively charged counter-ions. Simultaneously, there is a lack of HNCO identification in the study, hinting for the total conversion of HNCO into OCN$^-$ following the reaction HNCO + NH$_3$ → NH$_4^+$OCN$^-$, see \cite{Novozamsky,RAUNIER2003594,Fedoseev2015}. One may expect similar conversion of HCN into CN$^-$ to occur. Thus, acquisition of solid-state CN$^-$ spectroscopic data at various conditions can be a promising direction in the search of interstellar cyanides.


\end{document}